\documentclass[twocolumn,aps,prb,showpacs,groupedaddress]{revtex4}
\usepackage[T1]{fontenc}
\usepackage[latin1]{inputenc}
\usepackage{graphicx}
\usepackage{amssymb}

\begin{document}

\title{Phase diagram of geometric $d$-wave superconductor Josephson junctions}

\author{A. Gumann}
\author{N. Schopohl}
\affiliation{Institut f\"ur Theoretische Physik and Center for Collective Quantum Phenomena, Universit\"at T\"ubingen, Auf der Morgenstelle 14, D-72076 T\"ubingen, Germany}

\date{February 24, 2009}

\begin{abstract}
We show that a constriction-type Josephson junction realized by an epitactic thin film of a $d$-wave superconductor with an appropriate boundary geometry exhibits intrinsic phase differences between $0$ and $\pi$ depending on geometric parameters and temperature. Based on microscopic Eilenberger theory, we provide a general derivation of the relation between the change of the free energy of the junction and the current-phase relation. From the change of the free energy, we calculate phase diagrams and discuss transitions driven by geometric parameters and temperature.
\end{abstract}


\maketitle

\section{Introduction}

The change of the free energy of a Josephson junction (JJ) evoked by the variation of the phase determines the intrinsic phase difference in the unbiased ground state. Usually, the coupling energy between the electrodes of a JJ is positive and the current-phase relation is sinusoidal, corresponding to a vanishing intrinsic phase difference. For the peculiar case of negative coupling, however, intrinsic phase differences of $\pi$ are possible (see Refs. [\onlinecite{Gol02,Hil01}] and references therein).
\\
In the crossover regime between positive and negative coupling, higher harmonics dominate the current-phase relation. This behavior has been studied for different types of Josephson devices: tunneling and in particular grain boundary JJs involving $d$-wave superconductors~\cite{Yip01,Tan01,Il01}, controllable superconductor-normal metal-superconductor JJs (SNS)~\cite{Bas01}, superconductor-ferromagnet-superconductor JJs~\cite{Rad01,Sel01}, periodically alternating $0$-$\pi$ JJs~\cite{Min01,Buz01} and grain boundary JJs in noncentrosymmetric superconductors~\cite{Ini01}. The dominating higher harmonics close to the $0$-$\pi$ crossover lead to additional zeros of the current-phase relation. Whether or not these additional zeros are related to stable energetical minima, and accordingly to intrinsic phase differences of neither $0$ nor $\pi$, can only be decided by consideration of the free energy of the JJ.
\par
\begin{figure}
\includegraphics[width=0.9\columnwidth, keepaspectratio]{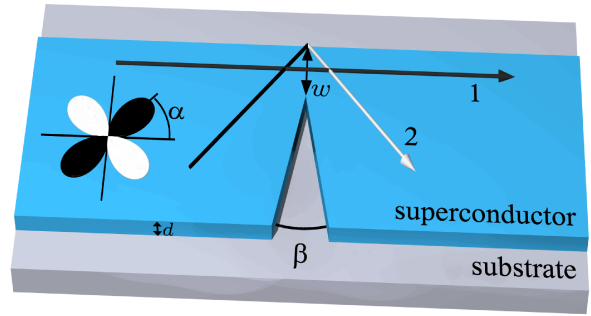}
\caption{\label{fig:gpiJJ}(Color online) Geometry defining the microbridge Josephson junction based on a $\hat{c}$-axis oriented epitactic thin film of a $d$-wave superconductor. A stripe of the superconducting material is narrowed down from one side by a wedge-shaped incision with the opening angle $\beta$. The width of the resulting microbridge-like junction is given by $w$. Two typical quasiparticle trajectories across the junction: 1-without reflection, 2-with reflection.}
\end{figure}

In the present work, we show that intrinsic phase differences in the full range $0\leq\gamma_0\leq\pi$ occur across $d$-wave superconductor microbridges. The geometry under consideration consists of a stripe of a $\hat{c}$-axis oriented epitactic thin film of a $d$-wave superconductor, which is narrowed down from one side by a wedge-shaped incision (Fig.~\ref{fig:gpiJJ}). Accordingly, the Josephson effect follows solely from the lateral constriction of the thin film and emerges if the width $w$ of the bridge is of the order of the coherence length $\xi_0$ of the superconducting material or below. Such a microbridge configuration is clearly distinct from a grain boundary, superconductor-isolator-superconductor or SNS tunneling JJ\cite{Bar01,Tan02}. It should be emphasized that in the geometry under consideration, Fig.~\ref{fig:gpiJJ}, there exists no grain boundary. According to the terminology introduced in Ref.~[\onlinecite{Lik01}], this microbridge configuration belongs to the weak link type JJs since the electrodes are not electrically separated by a tunneling barrier.

\par

In order to characterize the dc Josephson effect, we calculate current-phase relations based on microscopic Eilenberger theory. For the derivation of the intrinsic phase difference, we make use of the relation between the change of the free energy of the junction $\mathcal{E}(\gamma)-\mathcal{E}(0)$ and the current-phase relation $I(\gamma)$. We give a very general derivation of this relation which is valid for arbitrary structures exhibiting a current-phase relation in the full temperature range $0<T<T_c$ as well as in the presence of an external magnetic field. The intrinsic phase difference of the microbridge will be discussed in terms of phase diagrams, justified by the thoroughly derived change of the free energy.

\section{Basic Mechanism}

The intrinsic phase shift of the device is a direct consequence of the $d$-wave symmetry. If the width of the junction $w$ is large, quasiparticle trajectories without and with a reflection at the straight edge opposite to the wedge contribute to the total current across the junction (trajectories of type 1 and 2 in Fig.~\ref{fig:gpiJJ}). If the constriction is narrow enough, however, the dominant contribution to the total current stems from trajectories which get reflected (type 2). If the orientation of the $d$-wave is for example $\alpha=\pi/4$ (nodal surface), all reflected trajectories suffer a sign change of the pairing potential which leads to the formation of pronounced zero energy Andreev bound states at the surface~\cite{Hu01,Kas01,Loef01}. For $0<\alpha<\pi/4$, still a fraction of all trajectories contributes to the formation of zero energy Andreev bound states, which engender anomalous counterflowing quasiparticle surface currents~\cite{Fog01,Wal01} and also intrinsic phase shifts~\cite{Gum01}.

\section{Theory}

\subsection{Current-Phase Relation}

In order to calculate current-phase relations of the JJ, we employ microscopic Eilenberger theory~\cite{Eil01}. The selfconsistency equation which has to be solved for the pairing potential $\Delta(\mathbf{r},\mathbf{k}_F)$ reads
\begin{eqnarray}
\label{eq:gap}
\Delta(\mathbf{r},\mathbf{k}_F)
\!=\!
\int_{FS}\frac{d^2\mathbf{k}'_F}{(2\pi)^3}
\frac{[V_{pair}]_{\mathbf{k}_F,\mathbf{k}'_F}}{|\hbar \mathbf{v}'_F|}
2\pi k_B T
\!
\sum_{\varepsilon_{n'}>0}
\!
\frac{2a}{1+a\,b}
\end{eqnarray}
Here, $FS$ is the Fermi surface, $\mathbf{v}'_F=\mathbf{v}_F(\mathbf{k}'_F)$ is the Fermi velocity, $[V_{pair}]_{\mathbf{k}_F,\mathbf{k}'_F}$ is the pairing interaction matrix, $\varepsilon_n=(2n+1)\pi k_B T$ are fermionic Matsubara frequencies, and $a=a(\mathbf{r},\mathbf{k}'_F,\varepsilon_{n'})$ and $b=b(\mathbf{r},\mathbf{k}'_F,\varepsilon_{n'})$ are the Riccati amplitudes~\cite{Schop02,Schop01}. The selfconsistency equation for the current density $\mathbf{j}(\mathbf{r})$ is given by
\begin{eqnarray}
\label{eq:curr}
\mathbf{j}(\mathbf{r})
=
2e
\int_{FS}\frac{d^2\mathbf{k}'_F}{(2\pi)^3}
\frac{(-i) 2\pi k_B T}{|\hbar \mathbf{v}'_F|}
\sum_{\varepsilon_{n'}>0} \mathbf{v}'_F
\frac{1 - a\,b}{1 + a\,b}
\end{eqnarray}
The selfconsistency equations (\ref{eq:gap}), (\ref{eq:curr}) allow for the microscopic calculation of the current-phase relation~\cite{Gum01,Gum02}. Selfconsistent solutions guarantee current conservation ($\nabla\cdot\mathbf{j}(\mathbf{r})=0$), but can in general only be found numerically.

\subsection{Free Energy}

In order to derive the relation between the change of the free energy of the junction $\mathcal{E}(\gamma)-\mathcal{E}(0)$ and the current-phase relation $I(\gamma)$, we start from the Eilenberger functional~\cite{Eil01} for the free energy $\mathcal{E}(\Delta,\Delta^{\dagger},\mathbf{A};a,b)$, parametrized by the pairing potentials $\Delta$,  $\Delta^\dagger$, the vector potential $\mathbf{A}(\mathbf{r})$ and the Riccati amplitudes $a$, $b$:
\begin{widetext}
\begin{eqnarray}
\label{eq:eilfunc}
\mathcal{E}(\Delta,\Delta^{\dagger},\mathbf{A};a,b)
\,=\,
\int d^{3}r'
\left\{
\begin{array}{c}
\frac{1}{2\mu_0}
\mbox{rot}'\,\mathbf{A}\cdot\mbox{rot}'\,\mathbf{A}
-
\frac{1}{\mu_0}
\mathbf{B}_{ext}\cdot\mbox{rot}'\,\mathbf{A}
\vspace{0.2cm}
\\
+
\int_{FS}d^{2}\mathbf{k}'_{F}
\int_{FS}d^{2}\mathbf{k}_{F}''\,\Delta^{\dagger}(\mathbf{r}',\mathbf{k}_{F}')
[V_{pair}]^{-1}_{\mathbf{k}'_{F},\mathbf{k}_{F}''}\Delta(\mathbf{r}',\mathbf{k}_{F}'')
\vspace{0.2cm}
\\
-\int_{FS}\frac{d^{2}\mathbf{k}_{F}'}{(2\pi)^{3}}
\frac{2\pi k_B T}{|\hbar\mathbf{v}'_{F}|}
\sum\limits_{\varepsilon_{n'}>0}
\frac{2}{1+a\, b}
\left[\begin{array}{c}
\Delta^{\dagger}(\mathbf{r}',\mathbf{k}_{F}')\, a+\Delta(\mathbf{r}',\mathbf{k}_{F}')\, b
\\
+\left(
1-a\, b
\right)
\left(
\varepsilon_{n'}-i
\mathbf{v}'_{F}\cdot e\mathbf{A}
+\frac{1}{4}
\hbar
\mathbf{v}'_{F}\cdot\nabla'
\ln\frac{a}{b}
\right)
\end{array}
\right]
\end{array}
\right.
\end{eqnarray}
\end{widetext}
Here, $\mathbf{A}=\mathbf{A}(\mathbf{r}')$ is the total magnetic vector potential, $\mathbf{B}_{ext}=\mathbf{B}_{ext}(\mathbf{r}')$ is the external magnetic field, and $a=a(\mathbf{r}',\mathbf{k}'_F,\varepsilon_{n'})$ and $b=b(\mathbf{r}',\mathbf{k}'_F,\varepsilon_{n'})$ are the Riccati amplitudes.
\\
Consider a general variation of this functional:
\begin{eqnarray*}
d\mathcal{E}
\,=\,
\frac{\partial\mathcal{E}}{\partial\Delta}\, d\Delta
+\frac{\partial\mathcal{E}}{\partial\Delta^{\dagger}}\,d\Delta^{\dagger}
+\frac{\partial\mathcal{E}}{\partial\mathbf{A}}\, d\mathbf{A}
+\frac{\partial\mathcal{E}}{\partial a}\, da
+\frac{\partial\mathcal{E}}{\partial b}\, db
\end{eqnarray*}
The variation with respect to $\Delta^\dagger$ yields the selfconsistency equation for $\Delta(\mathbf{r},\mathbf{k}_F)$, Eq.~(\ref{eq:gap}), and the variation with respect to $\Delta$ a corresponding selfconsistency equation for $\Delta^\dagger(\mathbf{r},\mathbf{k}_F)$:
\begin{eqnarray}
\label{eq:gapconj}
\Delta^\dagger(\mathbf{r},\mathbf{k}_F)
\!=\!
\int_{FS}\frac{d^2\mathbf{k}'_F}{(2\pi)^3}
\frac{[V_{pair}]_{\mathbf{k}_F,\mathbf{k}'_F}}{|\hbar \mathbf{v}'_F|}
2\pi k_B T
\!
\sum_{\varepsilon_{n'}>0}
\!
\frac{2b}{1+a\,b}
\end{eqnarray}
The variation with respect to $b$ yields the Riccati differential equation for $a$ and vice versa~\cite{Schop02,Schop01}:
\begin{eqnarray}
\hbar
\mathbf{v}_F\cdot\nabla\, a
+ 2(\varepsilon_n - i \mathbf{v}_F\cdot e \mathbf{A} )\, a
+ \Delta^\dagger\, a^2
- \Delta
\,=\,
0
\\
\hbar
\mathbf{v}_F\cdot\nabla\, b
- 2(\varepsilon_n - i \mathbf{v}_F\cdot e \mathbf{A} )\, b
- \Delta\, b^2
+ \Delta^\dagger
\,=\,
0
\end{eqnarray}
Accordingly, all variations vanish in the case of a selfconsistent solution, i.e. at the stationary point of the functional, independent of the gauge.
\\
After making use of the selfconsistency equation for the currents, Eq.~(\ref{eq:curr}), and identification of the external currents via $\mbox{rot}\,\mathbf{B}_{ext}=\mu_0\,\mathbf{j}_{ext}$, we find for $\partial \mathcal{E} / \partial \mathbf{A}$:
\begin{eqnarray*}
d\mathcal{E}
\,=\,
\int d^{3}r'
\,
\bigg[
\Big(
\mathbf{j}(\mathbf{r}')
+
\mathbf{j}_{ext}(\mathbf{r}')
\Big)
-
\frac{1}{\mu_0}\,
\mbox{rot}\,\mbox{rot}\,\mathbf{A}(\mathbf{r}')
\bigg]
\,
d\mathbf{A}
\end{eqnarray*}

\begin{figure}[b]
\includegraphics[width=0.8\columnwidth, keepaspectratio]{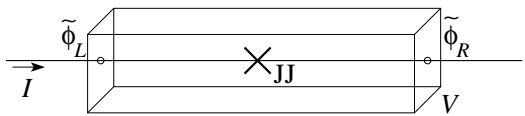}
\caption{\label{fig:intsketch}The volume $V$ for the integration of the free energy encloses the Josephson junction (JJ) through which a total current $I$ flows. Cross sections of the superconductor far from the junction are planes of constant gauge-invariant phase.}
\end{figure}

\noindent
If one integrates over all space, the variation of $\mathcal{E}$ with respect to $\mathbf{A}$ vanishes due to Maxwell's equation, $\mbox{rot}\,\mbox{rot}\,\mathbf{A}=\mu_0(\mathbf{j}+\mathbf{j}_{ext})$. Restricting the integration to a finite volume $V$ enclosing the junction, the external currents $\mathbf{j}_{ext}$ drop out. Now we carry out the gauge transformation
$ \{
\Delta \!\rightarrow\! \Delta e^{i\phi},
\Delta^\dagger \!\rightarrow\! \Delta^\dagger e^{-i\phi},
a \!\rightarrow\! a e^{i\phi},
b \!\rightarrow\! b e^{-i\phi},
\mathbf{A} \!\rightarrow\! \mathbf{A}+\frac{\hbar}{2 e}\nabla\phi
\} $, leading to
\begin{eqnarray}
d\mathcal{E}
\,=\,
\frac{\hbar}{2 e}
\int_V d^{3}r'
\,
\Big[
\,
\mathbf{j}(\mathbf{r}')
-
\frac{1}{\mu_0}\,
\mbox{rot}\,\mbox{rot}\,\mathbf{A}(\mathbf{r}')
\,
\Big]
\,
d(\nabla \tilde{\phi})
\end{eqnarray}
with the gauge-invariant phase
$\tilde{\phi}=\phi+\frac{2e}{\hbar}\int_\infty^\mathbf{r} d\mathbf{l}\cdot\mathbf{A}$.

\noindent
Considering the volume $V$ according to Fig.~\ref{fig:intsketch} and using basic vector calculus, it can be shown that the contribution of the vector potential vanishes. Integration by parts, exploitation of current conservation and application of Gauss's theorem then results in
\begin{eqnarray}
d\mathcal{E}
\,=\,
\frac{\hbar}{2e}
\int_{S=\partial V}
d\sigma'\,\mathbf{n}\cdot\mathbf{j}(\mathbf{r}')
\,
d\tilde{\phi}
\end{eqnarray}
Only the parts of the surface $S=\partial V$ where the current enters into or leaves the volume $V$ contribute. Since cross sections of the superconductor far from the junction are planes of constant gauge-invariant phase,
\begin{eqnarray}
d\mathcal{E}
\, = \, \frac{\hbar}{2e} \left[I\,d\tilde{\phi}_R - I\,d\tilde{\phi}_L \right]
\, = \, \frac{\hbar}{2e}\, I(\gamma)\, d\gamma
\end{eqnarray}
with the total current $I$ and the gauge-invariant phase difference $\gamma=\phi_R-\phi_L-\frac{2e}{\hbar}\int_{\mathbf{r}_L}^{\mathbf{r}_R} d\mathbf{l}\cdot\mathbf{A}$ and finally
\begin{eqnarray}
\label{eq:freeenergy}
\mathcal{E}\left(\gamma\right)-\mathcal{E}(0)
& = &
\frac{\hbar}{2e}\int_{0}^{\gamma}d\gamma'\, I\left(\gamma'\right)
\end{eqnarray}
Because of current conservation, the total current for the current-phase relation $I(\gamma)$ can be taken at any cross section of the superconductor $S_{sc}$:
\begin{eqnarray}
\label{eq:CPhiRriccati}
I(\gamma)
&\!=\!&
\int_{S_{sc}}\!
d\sigma' \mathbf{n}\cdot \mathbf{j}(\mathbf{r}')
\\
\nonumber
&\!=\!&
\int_{S_{sc}}\!
 d\sigma' \mathbf{n}
\!\cdot\!
2e\!
\int_{FS}\!
\frac{d^2\mathbf{k}'_F}{(2\pi)^3}
\frac{(-i) 2 \pi k_B T}{|\hbar \mathbf{v}'_F|}\!
\sum_{\varepsilon_{n'}>0}\! \mathbf{v}'_F \frac{1\!-\!a b}{1\!+\!a b}
\end{eqnarray}
\\
Quasiparticle bound states as well as the supercurrent contributions are included via the microscopic Riccati amplitudes $a$ and $b$.
\\
For the derivation of the result~(\ref{eq:freeenergy}), selfconsistency has been assumed. However, even if the current-phase relation used to evaluate Eq.~(\ref{eq:freeenergy}) has not been calculated selfconsistently, an upper bound for the change of the free energy follows. Eq.~(\ref{eq:freeenergy}) is valid at arbitrary temperature as well as in the presence of an external magnetic field.

\par

To the best of our knowledge, the derivation of Eq.~(\ref{eq:freeenergy}) given in the present work is the first microscopic derivation with general validity. Previous derivations were either based on thermodynamic reasoning and the application of the (second) Josephson relation $d\gamma/dt=2eV/\hbar$ with the voltage $V$ across the junction~\cite{Jos01} or were restricted to tunneling junctions~\cite{And01}. The derivation given in the present work does not depend on the actual realization of the JJ but is valid for arbitrary structures exhibiting a current-phase relation.

\begin{figure}
\includegraphics[width=0.99\columnwidth, keepaspectratio]{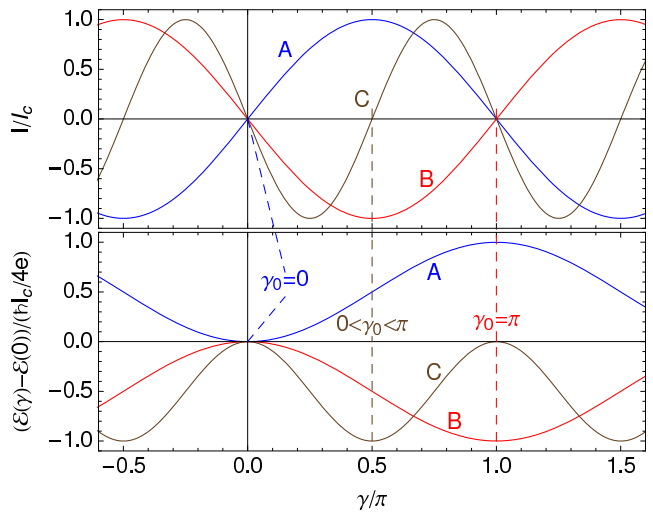}
\caption{\label{fig:FEsketch}(Color online) Three exemplary current-phase relations (upper panel) and the corresponding variations of the free energy (lower panel). \textsf{A}: $I=I_c\sin\gamma$, a normal or $0$-JJ with $\gamma_0=0$; \textsf{B}: $I=-I_c\sin\gamma$, a $\pi$-JJ with $\gamma_0=\pi$; \textsf{C}: $I=-I_c\sin(2\gamma)$, a so-called $\varphi$-JJ with $0<\gamma_0<\pi$.}
\end{figure}

\par
By examination of Eq.~(\ref{eq:freeenergy}), one finds that a zero of the current-phase relation $I(\gamma_0)=0$ with
$
d I(\gamma)/d\gamma
|_{\gamma=\gamma_0}
>0
$
corresponds to a stable local minimum of the free energy, and thus yields the intrinsic phase difference $\gamma_0$. If there exists no more than one nontrivial zero with $0<\gamma_0<\pi$, four cases can be distinguished: (1) $\gamma_0=0$ corresponds to a normal JJ; (2) $\gamma_0=\pi$ corresponds to a $\pi$-JJ. Finally, $0<\gamma_0<\pi$ and $I_c>0$ ($I_c<0$) corresponds to a so-called $\varphi$-JJ~\cite{Buz01,Gol01} with a positive (negative) critical current, where the critical current $I_c$ is defined as the absolute maximum of the current-phase relation.

\par

In Fig.~\ref{fig:FEsketch}, we sketch three exemplary current-phase relations and the corresponding variations of the free energy. In the case of a normal JJ with $\gamma_0=0$, the curvature of the variation of the free energy at $\gamma=0$ is positive, $d^2\mathcal{E(\gamma)}/d\gamma^2|_{\gamma=0}>0$ (see curves \textsf{A} in Fig.~\ref{fig:FEsketch}). For a $\pi$ JJ with $\gamma_0=\pi$, however, the curvature of the variation of the free energy at $\gamma=0$ is negative, $d^2\mathcal{E(\gamma)}/d\gamma^2|_{\gamma=0}<0$ (see curves \textsf{B} in Fig.~\ref{fig:FEsketch}). In the crossover regime with intermediate intrinsic phase differences $0<\gamma_0<\pi$, higher harmonics dominate, but still the curvature of the variation of the free energy at $\gamma=0$ is negative (see curves \textsf{C} in Fig.~\ref{fig:FEsketch}). Accordingly, in the case of a normal or $0$-JJ, the free energy of the junction firstly increases with increasing phase difference $\gamma>0$, whereas it firstly decreases in the case of $\pi$ and $\varphi$ JJs.

\section{Selfconsistent Solutions}

In this section, we present full two-dimensional selfconsistent solutions for the microbridge geometry depicted in Fig.~\ref{fig:gpiJJ}. Therefore, we numerically calculate a selfconsistent solution of the selfconsistency equation for the pairing potential $\Delta(\mathbf{r}, \mathbf{k}_F)$, Eq.~(\ref{eq:gap}). Based on this solution for the pairing potential, we numerically solve the equation for the current density $\mathbf{j}(\mathbf{r})$, Eq.~(\ref{eq:curr}).
\\
For the selfconsistent calculations, we assume a cylindrical Fermi surface with the cylinder axis aligned perpendicular to the film plane. The geometry used for the calculations spreads over an area of about $12.5\times 12.5\,\xi_0$ with the coherence length $\xi_0=\hbar v_F /(\pi \Delta_\infty(T=0))$. Specular boundary conditions lead to $\hat{\mathbf{n}}\cdot \mathbf{j}=0$ with the surface normal $\hat{\mathbf{n}}$ at all surfaces of the geometry. For the left and right end of the geometry depicted in Fig.~\ref{fig:gpiJJ}, periodic boundary conditions have been used. Selfconsistency automatically guarantees current conservation, $\nabla \cdot \mathbf{j}=0$. Details of the selfconsistent calculations have been published in a previous work on $s$-wave superconducting microbridges\cite{Gum02}.

\begin{figure}
\includegraphics[width=0.99\columnwidth, keepaspectratio]{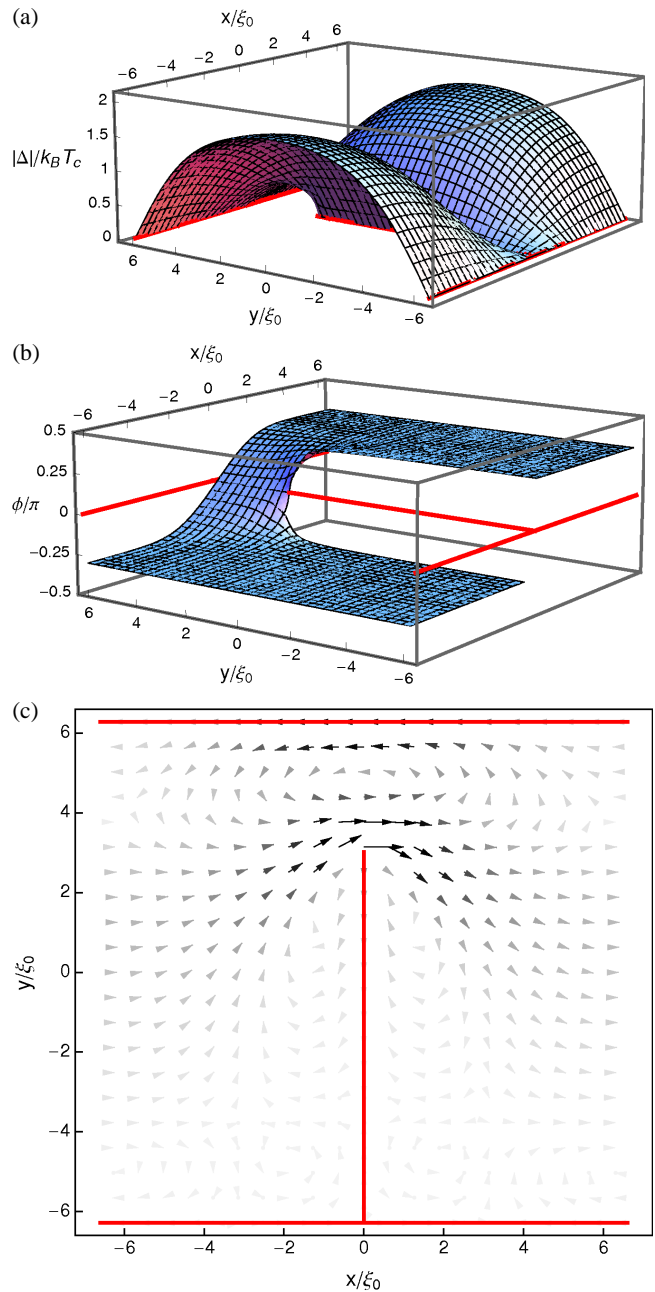}
\caption{\label{fig:selfconsT05}(Color online) Amplitude (a) and phase (b) of the pairing potential as well as the corresponding current density (c) at the critical current for a $d$-wave orientation angle of $\alpha=\pi/4$, an opening angle of the wedge of $\beta=0$ and a width of $w=3.14\,\xi_0$ at a temperature of $T=0.5\,T_c$. The boundary geometry is indicated by the thick (red) lines. In (c), the shading is proportional to the current density.}
\end{figure}

\par

In Fig.~\ref{fig:selfconsT05}, we present selfconsistent configurations for the amplitude and the phase of the pairing potential as well as for the corresponding current density. For this figure, an orientation angle of the $d$-wave of $\alpha=\pi/4$ has been used because, in this case, the effect of the $d$-wave symmetry is most pronounced. The opening angle of the wedge has been chosen to be $\beta=0$ and the width of the microbridge $w=3.14\,\xi_0$. An intermediate temperature of $T=0.5\,T_c$ has been used.
\\
For the orientation angle $\alpha=\pi/4$, the $d$-wave symmetry leads to a suppression of the amplitude of the pairing potential at all surfaces of the rectangular geometry (see Fig.~\ref{fig:selfconsT05} (a)). The phase difference $\gamma$ across the microbridge which has been used for this figure corresponds to the critical current. At $T=0.5\,T_c$, the phase of the pairing potential monotonically increases from $-\gamma/2$ at the left boundary of the geometry to $+\gamma/2$ at the right boundary (see Fig.~\ref{fig:selfconsT05} (b)). The corresponding current distribution in Fig.~\ref{fig:selfconsT05} (c) exhibits contributions which flow along the gradient of the phase (from left to right, in positive direction) as well as backflowing surface currents\cite{Fog01,Wal01}. These backflowing surface currents are directly related to the $d$-wave symmetry and they are carried by Andreev bound states which exist at surfaces of $d$-wave superconductors with orientation angles $\alpha\neq0$\cite{Hu01,Kas01,Loef01}. However, if one integrates the current density shown in Fig.~\ref{fig:selfconsT05} (c) over a cross section of the geometry, a positive total current follows.

\begin{figure}
\includegraphics[width=0.99\columnwidth, keepaspectratio]{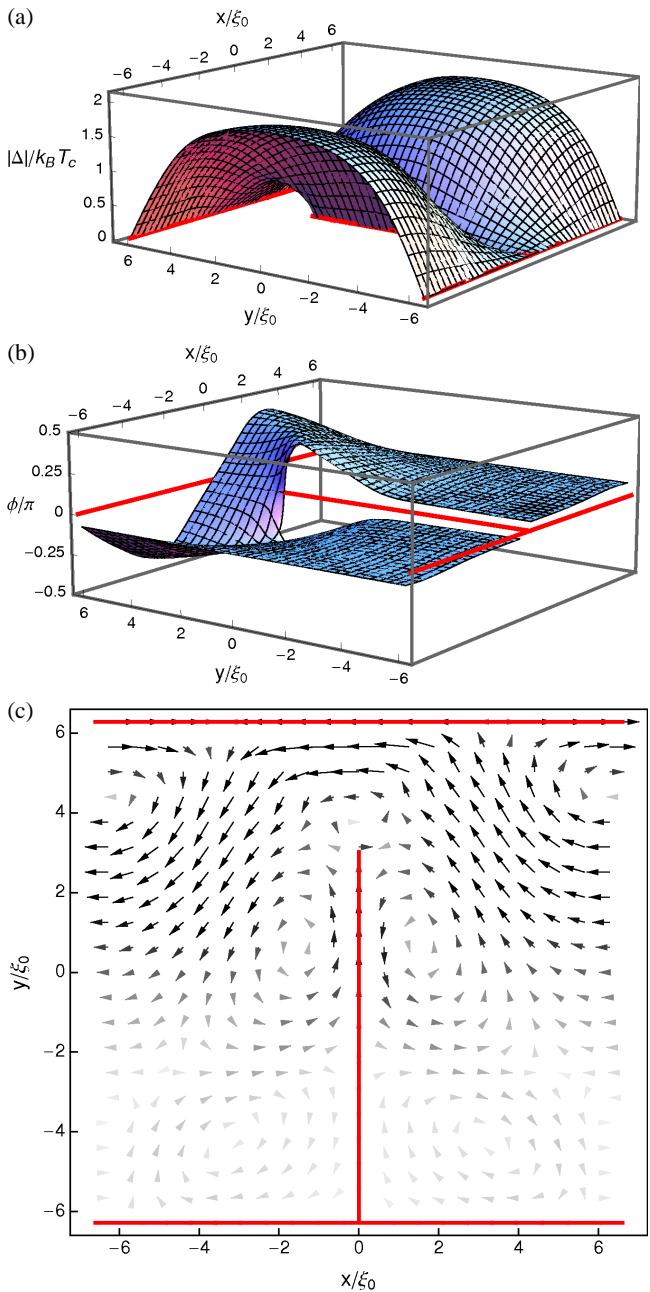}
\caption{\label{fig:selfconsT01}(Color online) Amplitude (a) and phase (b) of the pairing potential as well as the corresponding current density (c) at a temperature of $T=0.1\,T_c$. Other parameters and presentation corresponding to Fig.~\ref{fig:selfconsT01}.}
\end{figure}

\par

In Fig.~\ref{fig:selfconsT01}, we present selfconsistent solutions corresponding to those in Fig.~\ref{fig:selfconsT05}, but for a lower temperature of $T=0.1\,T_c$. From Fig.~\ref{fig:selfconsT01} (a), it is obvious that the amplitude of the pairing potential hardly changes between $T=0.5\,T_c$ and $T=0.1\,T_c$. However, the configuration of the phase of the pairing potential changes completely (see Fig.~\ref{fig:selfconsT01} (b)). With decreasing temperature, the influence of surface Andreev bound states strongly increases, which leads to dominating backflowing surface currents (Fig.~\ref{fig:selfconsT01} (c)). Accordingly, the phase exhibits a nonmonotonic variation with a phase shift (Fig.~\ref{fig:selfconsT01} (b)). At $T=0.1\,T_c$, integrating the current density over a cross section of the geometry yields a negative total current.

\par

From the selfconsistent configurations presented in Figs.~\ref{fig:selfconsT05} and \ref{fig:selfconsT01}, it follows that the behavior of the microbridge strongly depends on the temperature $T$. With decreasing temperature, the influence of surface Andreev bound states increases and, accordingly, the relative weight of the backflowing surface currents. An increasing importance of surface Andreev bound states with decreasing temperature has also been found in a previous work on the influence of the surface Andreev bound states on the Bean-Livingston barrier~\cite{Ini03}. The backflowing surface currents dominate at low temperatures $T$ as well as for small widths $w$ of the junction. This leads to negative values of the current-phase relations~\cite{Gum03} and even to negative critical currents~\cite{Gum01}. As follows from Eq.~(\ref{eq:freeenergy}) which links the current-phase relation and the variation of the free energy, these negative currents are related to finite intrinsic phase differences $0<\gamma_0\leq\pi$. The intrinsic phase difference $\gamma_0$ determines the state of the microbridge JJ, which can be either a normal or $0$-JJ, a $\varphi$-JJ or a $\pi$-JJ.

\par

The two-dimensional selfconsistent solutions shown in Figs.~\ref{fig:selfconsT05} and \ref{fig:selfconsT01} each are representative configurations for one set of the parameters $T$, $\alpha$, $\beta$ and $w$ and for a fixed value of the phase difference $\gamma$. Fig.~\ref{fig:selfconsT05} for $T=0.5\,T_c$ corresponds to a normal or $0$-JJ whereas Fig.~\ref{fig:selfconsT01} corresponds to a $\pi$-JJ. In order to obtain current-phase relations and, accordingly, intrinsic phase differences for diverse combinations of the relevant parameters, the full two-dimensional calculations have to be repeated many times.

\par

Finally, it should be noted that the selfconsistent configurations shown in Figs.~\ref{fig:selfconsT05} and \ref{fig:selfconsT01} do not change substantially if the orientation of the $d$-wave deviates from $\alpha=\pi/4$. Small deviations do not lead to an abrupt disappearance of the backflowing surface currents. Similarly, opening angles of the wedge other than $\beta=0$ do not lead to an abrupt disappearance of the backflowing surface currents and do not substantially change the current configurations as presented in Figs.~\ref{fig:selfconsT05} and \ref{fig:selfconsT01}.

\section{Step Model}

In order to calculate complete phase diagrams of the microbridge, the selfconsistent calculations presented in the last section would have to be repeated for arbitrary combinations of the relevant parameters: temperature $T$, width of the junction $w$, orientation angle of the $d$-wave $\alpha$, opening angle of the wedge $\beta$ and, finally, in order to obtain current-phase relations, phase differences across the junction in the range $0\leq\gamma\leq\pi$. Unfortunately, the numerical costs of the full two-dimensional selfconsistent calculations inhibit the selfconsistent calculation of complete phase diagrams. Therefore, in the following, we employ a non-selfconsistent step model for the pairing potential $\Delta(\mathbf{r},\mathbf{k}_F)$ in order to calculate current-phase relations and the according variations of the free energy for the microbridge. Nevertheless, based on the microscopic derivation of the relation between the current-phase relation and the variation of the free energy, Eq.~(\ref{eq:freeenergy}), we know that a non-selfconsistent calculation provides an upper bound for the variation of the free energy. A detailed comparison of the full two-dimensional selfconsistent calculations and the step model will be discussed subsequently.

\par

\begin{figure}
\includegraphics[width=0.99\columnwidth, keepaspectratio]{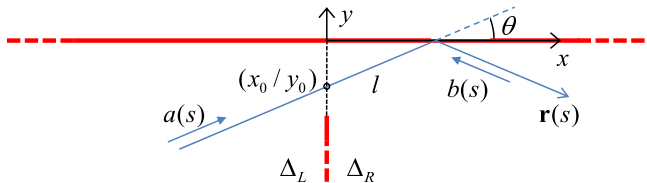}
\caption{\label{fig:stepsketch}(Color online) Boundary geometry for the step model, Eq.~(\ref{eq:step}), together with an exemplary quasiparticle trajectory $\mathbf{r}(s)$ and the Riccati amplitudes $a(s)$, $b(s)$ corresponding to Eqs.~(\ref{eq:stepab0}).}
\end{figure}

We assume a cylindrical Fermi surface with the cylinder axis aligned perpendicular to the film plane. Accordingly, $\mathbf{v}_F=v_F(\hat{\mathbf{x}}\cos \theta + \hat{\mathbf{y}}\cos \theta )$ with $\hat{\mathbf{x}},\hat{\mathbf{y}}$ being unit vectors in the film plane and $\theta$ being the polar angle. Thus, in the case of $d$-wave pairing, $V_{\mathbf{k}_F,\mathbf{k}'_F}=V\cos(2\theta-2\alpha)\cos(2\theta'-2\alpha)$. The step model corresponds to an opening angle of the wedge of $\beta=0$ and assumes a step-like variation of the phase of the pairing potential, whereas its amplitude is taken to be constant:
\begin{eqnarray}
\label{eq:step}
\Delta_{L,R}(\mathbf{r},\mathbf{k}_F)
&=&
\Delta_\infty(T)\cos(2\theta-2\alpha)e^{\mp i \gamma/2}
\end{eqnarray}
Here, the indices $L,R$ label the left and right side of the junction and $\Delta_\infty(T)$ is the temperature-dependent amplitude of the pairing potential in the bulk. The step model (\ref{eq:step}) has to be solved taking into account the boundary geometry defining the microbridge JJ (see Fig.~\ref{fig:stepsketch}).
\\
In order to find the current density~(\ref{eq:CPhiRriccati}) at the cross section of the constriction, we solve the Riccati equations along trajectories $\mathbf{r}(s)=(x_0=0,y_0)+s\,(\cos\theta, \sin\theta)$, see Fig.~\ref{fig:stepsketch}. Introducing
\begin{eqnarray*}
\Omega(\theta)
&=&
\sqrt{\varepsilon_n^2+|\Delta_\infty(T)\cos(2\theta-2\alpha)|^2}
\\
a_{L,R}(\theta)
&=&
b_{L,R}^\dagger(\theta)
\,\,=\,\,
\frac
{\Delta_{L,R}(\theta)}
{\varepsilon_n+\Omega(\theta)}
\end{eqnarray*}
as well as $\eta(\theta)=2\Omega(\theta)/(\hbar v_F)$ and $l=|y_0/\sin\theta|$, we find for $0<\theta<\pi/2$:
\begin{eqnarray}
\label{eq:stepab0}
a(s\!=\!0)
&\!=\!&
a_L(\theta)
\\
\nonumber
b(s\!=\!0)
&\!=\!&
b_R(\theta)
+
\frac{1}
{
\frac{e^{\eta(\theta)l}}
{
b_R(2\pi\!-\!\theta)
-
b_R(\theta)}
\!+\!
\frac{\Delta_R(\theta)
}{2\Omega(\theta)}
(e^{\eta(\theta)l}-1
)
}
\end{eqnarray}
Results for $\pi/2<\theta<2\pi$ follow accordingly.
\\
Based on Eqs.~(\ref{eq:stepab0}), the total current (\ref{eq:CPhiRriccati}) can be calculated by integrating the current density over the cross section of the microbridge JJ, i.e. along the negative $y$-axis of the geometry depicted in Fig.~\ref{fig:stepsketch}. The integral over the cross section has to be taken from $y=0$ to the width of the junction at $y=-w$.

\section{Results And Discussion}

\begin{figure}
\includegraphics[width=0.99\columnwidth, keepaspectratio]{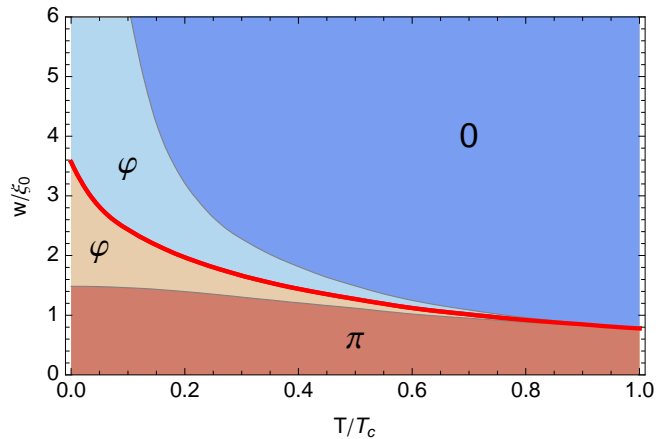}
\caption{\label{fig:PDtot}(Color online) Phase diagram of the geometric Josephson junction shown in Fig.~\ref{fig:gpiJJ}. The thick (red) line separates regions of positive and negative critical current (above and below). For this figure, $\alpha=\pi/4$.}
\end{figure}

Based on the calculation of current-phase relations and critical currents $I_c$ from the step model~(\ref{eq:step}), we show in Fig.~\ref{fig:PDtot} the phase diagram of the geometric microbridge JJ from Fig.~\ref{fig:gpiJJ} for a fixed $d$-wave orientation angle $\alpha=\pi/4$. From the phase diagram, it follows that if the width of the junction is smaller than a critical value $w_c$, the critical current is negative. For $T\rightarrow0$, we find a value of about $w_c\approx 3.6\,\xi_0$. With increasing temperature, the critical width decreases to about $w_c\approx 0.78\,\xi_0$ near $T_c$. Near $T=T_c$, the current-phase relations assume the asymptotic forms $I= I_c\sin\gamma$ for $w>w_c$ and $I=-I_c\sin\gamma$ for $w<w_c$, respectively, and only the $0$ and the $\pi$ state occur. With decreasing temperature, higher harmonics of the current-phase relations become more important and the $\varphi$ state appears in the vicinity of the $0$-$\pi$ transition. At low temperatures, the $\varphi$ state extends to widths $w$ much larger than the critical width $w_c$ which separates $I_c<0$ from $I_c>0$.

\begin{figure}
\includegraphics[width=0.99\columnwidth, keepaspectratio]{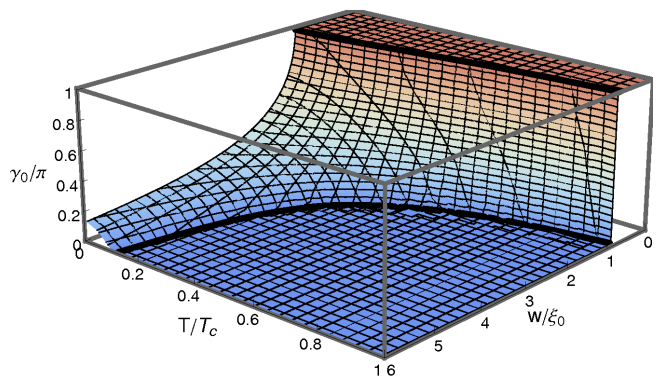}
\caption{\label{fig:PDtot_pd}(Color online) Intrinsic phase difference $\gamma_0$ corresponding to the phase diagram in Fig.~\ref{fig:PDtot}. For this figure, $\alpha=\pi/4$.}
\end{figure}

\par
In the limit $w\rightarrow 0$, all trajectories traveling through the junction suffer a reflection. Accordingly, this situation can be considered as a $\pi$ point contact, the complementary configuration to a normal point contact~\cite{Kul01}. The limit $w\rightarrow 0$ implies $l\rightarrow 0$ and Eqs.~(\ref{eq:stepab0}) become particularly simple:
\begin{eqnarray}
a(s\!=\!0)
\,=\,
a_L(\theta)
&\hspace{0.5cm}&
b(s\!=\!0)
\,=\,
b_R(2\pi-\theta)
\end{eqnarray}
In this case, the current-phase relations of the $d$-wave point contact are being reproduced, but with an intrinsic phase shift of $\gamma_0=\pi$. From the phase diagram in Fig.~\ref{fig:PDtot}, one finds that the $\pi$ point contact exists at all temperatures $0<T<T_c$.

\begin{figure}
\includegraphics[width=0.99\columnwidth, keepaspectratio]{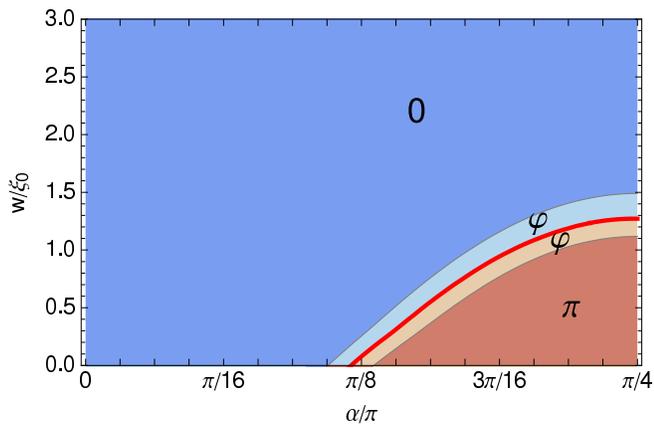}
\caption{\label{fig:PDvaralpha}(Color online) Phase diagram for the variation of the $d$-wave orientation angle $\alpha$. The thick (red) line separates regions of positive and negative critical current (above and below). For this figure, $T=0.5\,T_c$.}
\end{figure}

\begin{figure}
\includegraphics[width=0.99\columnwidth, keepaspectratio]{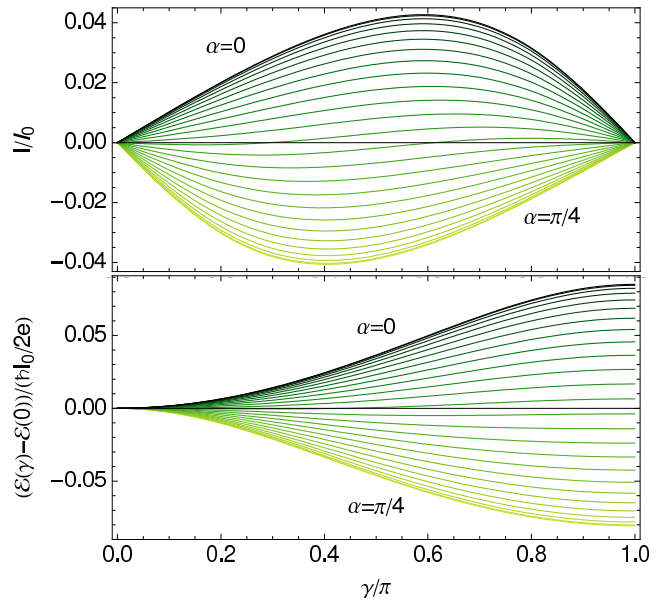}
\caption{\label{fig:PDvaralpha_cphir}(Color online) Current-phase relations (upper panel) and the corresponding variations of the free energy (lower panel) for angles $\alpha$ from $\alpha=0$ to $\alpha=\pi/4$ (as indicated) in steps of $\pi/100$ for $w=0.1\,\xi_0$. The scale for the currents is given by $I_0=\pi e N(0) v_F k_B T_c \xi_0 d$. For this figure, $T=0.5\,T_c$.}
\end{figure}

\par
In Fig.~\ref{fig:PDtot_pd}, we show the intrinsic phase difference $\gamma_0$ corresponding to the phase diagram in Fig.~\ref{fig:PDtot}. The disappearance of the $\varphi$ state near $T=T_c$ becomes apparent as a discontinuous transition from $\gamma_0=0$ to $\gamma_0=\pi$. At lower temperatures, however, a continuous transition arises.

\par
In Fig.~\ref{fig:PDvaralpha}, we plot the phase diagram for a fixed temperature of $T=0.5\,T_c$, focusing on the variation of the orientation of the $d$-wave $\alpha$. Starting from the ideal orientation $\alpha=\pi/4$ for the occurrence of the $\pi$ state, we find that the critical width $w_c$ decreases when the $d$-wave is being rotated. However, small deviations from $\alpha=\pi/4$ do not lead to an abrupt disappearance of the $\pi$ or the $\varphi$ state which is important for the experimental realization.

\par
In Fig.~\ref{fig:PDvaralpha_cphir}, we show current-phase relations and the corresponding variations of the free energy for $d$-wave orientation angles $\alpha$ from $\alpha=0$ to $\alpha=\pi/4$. The $\pi$ state is apparent for $\alpha$ close to $\pi/4$ since negative currents occur for all $\gamma$. Accordingly, close to $\alpha=\pi/4$, the free energy decreases with increasing $\gamma$. With decreasing $\alpha$, a transition to the $0$ state occurs, with positive currents and an increasing free energy for all $\gamma$. Close to the very transition, a $\varphi$ region with an additional zero of the current-phase relation occurs. Since the gradient of the current-phase relation at this additional zero is positive, it corresponds to a stable energetical minimum and an intermediate intrinsic phase difference $0<\gamma_0<\pi$.

\par
In the present work, we use the step model~(\ref{eq:step}) in order to calculate phase diagrams of the geometric microbridge Josephson junction depicted in Fig.~\ref{fig:gpiJJ}. From the microscopic derivation of the relation between the current-phase relation and the variation of the free energy, we know that an upper bound for the free energy follows if a non-selfconsistent model for the pairing potential is employed. Based on a detailed comparison of current-phase relations from the step model and from full two-dimensional selfconsistent solutions~\cite{Gum03}, we do not expect selfconsistency to qualitatively alter the general properties of the device as described here. We rather find that the critical width $w_c$ which marks the transition to negative critical currents is underestimated in the scope of the non-selfconsistent step model. It should be noted that, according to the selfconsistent calculations, the opening angle of the wedge $\beta$ hardly influences the current-phase relations. However, small opening angles of the wedge could possibly lead to an increased capacitive coupling of the electrodes. As long as the width $w$ is in the range of several $\xi_0$ or below, the step model proves to be a useful approximation. Since microscopic surface roughness does not suppress surface Andreev bound states~\cite{Fog01,Ini02}, the reported intrinsic phase differences are expected to be robust features.

\section{Conclusion}

In the present work, we provide a microscopic derivation of the relation between the variation of the free energy and the current-phase relation with very general validity. Based on a comparison with full two-dimensional selfconsistent solutions, we use a step model for the pairing potential in order to calculate current-phase relations and intrinsic phase differences for the geometric microbridge Josephson junction depicted in Fig.~\ref{fig:gpiJJ}. The calculation of intrinsic phase differences is used to access phase diagrams of the Josephson junction, justified by the relation to the variation of the free energy.

\begin{figure}
\includegraphics[width=0.95\columnwidth, keepaspectratio]{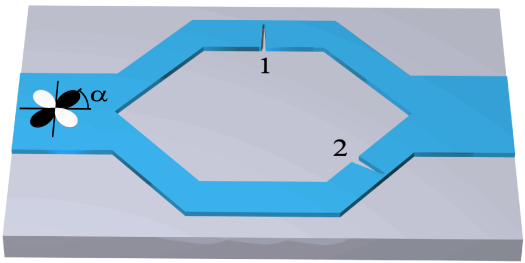}
\caption{\label{fig:squid_D6}(Color online) Dc superconducting quantum interference device (dc SQUID) geometry consisting of two geometric microbridge Josephson junction according to Fig.~\ref{fig:gpiJJ}. For $\alpha=\pi/4$, the geometric microbridge Josephson junction labeled 1 can be in the $0$, $\varphi$ or $\pi$ state depending on its width and on the temperature whereas the Josephson junction labeled 2 always is in the normal or $0$ state\cite{Kem01}.}
\end{figure}

\par

From the phase diagrams, we conclude that the structure sizes required for the experimental realization of the $\pi$ and in particular of the $\varphi$ state in cuprate high-temperature superconductor microbridges are within reach of modern fabrication technology. Because of the larger coherence length, electron-doped materials are especially promising~\cite{Fab01}. To test our predictions we suggest an interference experiment with a dc superconducting quantum interference device (dc SQUID) consisting of two microbridge JJs with $\alpha_1=\pi/4$ and $\alpha_2=0$, respectively (see Fig.~\ref{fig:squid_D6}). Geometry- and temperature-dependent intrinsic phase differences according to the phase diagram will show up as shifts of the corresponding flux-dependent interference pattern. The experimental verification of the intrinsic phase differences would at the same time imply a direct confirmation of the anomalous counterflowing quasiparticle surface currents which are a unique and intriguing fingerprint of $d$-wave pairing symmetry.

\begin{acknowledgments}
We acknowledge useful discussions with C.~Iniotakis, T.~Dahm, B.~Gro\ss, M.~Kemmler, R.~Kleiner and D.~Koelle.
\end{acknowledgments}

\end{document}